\documentclass[final,5p,times, twocolumn]{elsarticle_accept}

\date{4/22/2014}

\usepackage{graphics}
\usepackage{amssymb}
\newcounter{bla}

\journal{Computer Physics Communications}

\begin{document}

\begin{frontmatter}

\title{KMCLib: A general framework for lattice
  kinetic Monte Carlo (KMC) simulations}

\author[a]{Mikael Leetmaa\corref{author}}
\author[a,b]{Natalia V.\ Skorodumova}

\cortext[author] {Corresponding author.\\\textit{E-mail address:} leetmaa@kth.se}
\address[a]{Multiscale Materials Modelling, Materials Science and Engineering, School of Industrial Engineering and Management, KTH - Royal Institute of Technology, Brinellv\"{a}gen 23, 100 44 Stockholm, Sweden }
\address[b]{Department of Physics and Astronomy, Uppsala University, Box 516, 751 20 Uppsala, Sweden}

\begin{abstract}
KMCLib is a general framework for {\it lattice} kinetic Monte Carlo (KMC)
simulations.
The program can handle simulations of the diffusion and reaction
of millions of particles in one, two, or three dimensions, and is
designed to be easily
 extended and customized by the user to allow for the development of
 complex custom KMC models for specific systems
 without having to modify the core functionality of the program.
Analysis modules and on-the-fly elementary step diffusion rate
calculations can be implemented as plugins following a well-defined
API. The plugin modules are loosely coupled to the core KMCLib
program via the Python scripting language.
KMCLib is written as a Python module with a backend C++
library. After initial compilation of the backend library KMCLib is
used as a Python module; input to the program is given as a
Python script executed using a standard Python interpreter.
We give a detailed description of the features and implementation of
the code and demonstrate its scaling behavior and parallel performance with
a simple one-dimensional A-B-C lattice KMC model and a more complex
three-dimensional lattice KMC model of oxygen-vacancy diffusion in a
fluorite structured metal oxide.
KMCLib can keep track of individual particle movements and includes
tools for mean square displacement analysis,
and is therefore particularly well suited for studying diffusion
processes at surfaces and in solids.

\end{abstract}


\end{frontmatter}

\noindent
{\bf PROGRAM SUMMARY}\\

\begin{small}
\noindent
{\em Program Title:}     KMCLib                               \\
{\em Journal Reference:}                                      \\
{\em Catalogue identifier:}                                   \\
{\em Licensing provisions:} GPLv3 \\
{\em Programming language:} Python and C++                 \\
{\em Computer:} Any computer that can run a C++ compiler and a Python
interpreter \\
{\em Operating system:} Tested on Ubuntu 12.4 LTS, CentOS release 5.9,
Mac OSX 10.5.8 and Mac OSX 10.8.2, but should run on any
system that can have a C++ compiler, MPI and a Python interpreter.\\
{\em RAM:} From a few megabytes to several
gigabytes depending on input parameters and the size of the system to simulate.\\
{\em Number of processors used:} From one to hundreds of processors
depending on the on the type of input and simulation. \\
{\em Supplementary material:} The full documentation of the program is distributed with the code and
can also be found at {\tt http://www.github.com/leetmaa/KMCLib/manual}
{\em Classification:}   4.13, 16.13                                      \\
{\em External routines/libraries:} KMCLib uses and external
Mersenne Twister pseudo random number generator that is included in
the code. A Python 2.7 interpreter and a standard C++ runtime library
is needed to run the serial version of the code. For running the
parallel version an MPI implementation is needed, such as {\it
  e.g.\ }MPICH from {\tt http://www.mpich.org} or OpenMPI
from {\tt http://www.open-mpi.org}. SWIG (obtainable from {\tt
  http://www.swig.org/}) and CMake
(obtainable from {\tt http://www.cmake.org/}) are needed for building the
backend module, Sphinx (obtainable from {\tt http://sphinx-doc.org})
for building the documentation and CPPUNIT (obtainable from
{\tt http://sourceforge.net/projects/cppunit/}) for building the C++
unit tests. \\
{\em Nature of problem:}\\
Atomic scale simulation of slowly evolving dynamics is a great
challenge in many areas of computational materials science and
catalysis. When the rare-events dynamics of interest is orders of
magnitude slower than the typical atomic vibrational frequencies
a straight-forward propagation of the equations of motions for the
particles in the simulation can not reach time scales of
relevance for modeling the slow dynamics.
   \\
{\em Solution method:}\\
KMCLib provides an implementation of the kinetic Monte Carlo (KMC)
method that solves the slow dynamics problem by utilizing the separation
of time scales between fast vibrational motion and
the slowly evolving rare-events dynamics. Only the latter is treated
explicitly and the system is simulated as jumping between fully
equilibrated local energy minima on the slow-dynamics potential
energy surface.
   \\
{\em Restrictions:}\\
KMCLib implements the {\it lattice} KMC method and is as such
restricted to geometries that can be expressed on a grid in space.
   \\
{\em Unusual features:}\\
KMCLib has been designed to be easily customized, to allow for
user-defined functionality and integration with other codes.
The user can define her own on-the-fly rate calculator via a Python
API, so that site-specific elementary process rates, or rates
depending on long-range interactions or complex geometrical features
can easily be included. KMCLib also allows for on-the-fly analysis with
user-defined analysis modules.
KMCLib can keep track of individual particle movements and includes
tools for mean square displacement analysis,
and is therefore particularly well suited for studying diffusion
processes at surfaces and in solids.
   \\
{\em Additional comments:}\\
The program can be obtained as a git repository from {\tt http://www.github.com/leetmaa/KMCLib}
   \\
{\em Running time:}\\
From a few seconds to several days depending on the type of simulation
and input parameters.
   \\
\end{small}

\vspace{14mm}

\section{Introduction}
\label{Introduction}

Atomic scale simulations of slowly evolving dynamics pose a great
challenge in many areas of computational materials science and catalysis.
The dynamics of processes such as chemical reactions on surfaces
and slow diffusion events in solids often involve time-scales that are
impossible to access with conventional molecular dynamics (MD) simulations.
When atomic resolution is demanded to properly describe the
investigated physical system, but the time-scale of the
events of interest renders MD simulations
unsuitable, some other method working efficiently with atomic
resolution on longer time scales must be employed.
There exists today a range of accelerated dynamics methods handling the
dynamics in one way or another to allow for the rare events to occur
much more frequently in the simulation \cite{AcceleratedDynamicsReview}.
 Kinetic Monte Carlo (KMC) \cite{OriginalKMC1}, \cite{OriginalKMC2}
is one such method.

The idea behind KMC is to utilize the separation of time
scales between on the one hand the fast molecular vibrations, and on the
other hand the rare events of interest evolving the slow dynamics.
Only the latter is explicitly simulated, while the former is treated as
thermally equilibrated at the time one of the rare events takes place.
This is theoretically justified when the
separation of the vibrations and rare events time scales is large
\cite{TheoreticalDynamicMC}, \cite{KMCOverview}. For many
systems of high scientific and technological interest this condition
is known to be well satisfied, paving the ground for the
popularity of different flavors of the KMC method \cite{KMCOverview},
\cite{aKMC2}, \cite{AS-KMC}, \cite{BioKMC}, \cite{SSAReview},
with common applications
ranging from biological systems \cite{BioKMC} and heterogeneous
catalysis \cite{CatalysisKMC} to the electrochemistry in fuel-cells
\cite{FuelCellKMC} and atomic transport mechanisms in alloys \cite{AlloyKMC}.

We will in this work focus
on the {\it lattice} KMC method, a particularly efficient
form of KMC for simulating ordered -- as opposed to amorphous --
systems, where the geometry of the simulated system can be
conveniently described on a regular grid in space \cite{KMCOverview}.
KMCLib implements the {\it lattice} KMC algorithm in the form
of the algorithm known as the variable step size method (VSSM) or the
{\it n-fold} way\cite{OriginalKMC1}, \cite{OriginalKMC2}, \cite{KMCOverview}.
Although several other methods are possible \cite{CARLOS} the VSSM
methods is theoretically well founded
\cite{Foundation} and particularly well suited for an
efficient implementation for simulations of large systems.

Despite its maturity and the popularity of the approach
there has been surprisingly few publicly available computer programs
implementing the {\it lattice} KMC method \cite{CARLOS}, \cite{CARLOS_web},
\cite{SPPARKS}, \cite{SPPARKS_web}. Although this situation is
improving \cite{KMOS}, \cite{KMOS_web}, \cite{ZACROS}, most
available codes are still particularly geared towards surface studies
\cite{CARLOS}, \cite{CARLOS_web}, \cite{ZACROS}, or closed source
\cite{CARLOS_web} preventing straightforward customization.
Due to this shortage of publicly available open sourced software
together with the relative
simplicity of the algorithm and the need for significant method
customizations in many cases, is has not been uncommon that a whole new KMC
code gets written for each new set of physical problems to solve. Many
research groups each have their own custom code aimed at
simulating a particular type of systems, and it is not uncommon that
these codes never make it to the public.
We hope to reverse this trend by providing a fast and efficient
general framework for {\it lattice} KMC simulations that is easy to
customize and to integrate with other software.
Our hope is that
KMCLib will be a useful platform for
implementing custom KMC models, meeting the highest expectations
for flexibility, robustness and efficiency.

KMCLib consists of a highly optimized C++ core program with a
well-defined front-end API in Python, and is after compilation of the
C++ code used as a Python module. Code extensions can be written
in Python with no need to recompile the underlying C++ code.
We have introduced the possibility for using
custom rate calculations on-the-fly during the simulation, via a Python
interface, that facilitates usage of any external code for calculating
elementary process rates. This makes it possible to include
{\it e.g.\ }long-range
interactions or site and geometry dependent process-rates
in the simulations, without modifying the main program.
A plugin API for analysis on-the-fly during the simulation has also
been specified to allow for user-defined analysis tools to be
integrated with the code. The code is also equipped with an
implementation of a recent mean square displacement
(MDS) algorithm for non-equidistant time step KMC simulations
\cite{MSDAlgorithm} to
facilitate diffusion studies with the code.

This paper describes the main simulation algorithms implemented
in KMCLib with and without the custom rate calculation modifications.
The code is then described in more detail with focus on technical
solutions.
The scaling behavior of the implementation is demonstrated using a
simple one-dimensional A-B-C model as well as with a more complex
three-dimensional fluorite structure metal oxide vacancy-diffusion
model. Parallel performance is finally discussed and demonstrated.
The KMC formalism itself will not be described in detail since it
is well-known and has been described elsewhere in the literature
(see e.g.\ \cite{OriginalKMC1}, \cite{OriginalKMC2}, \cite{KMCOverview}, \cite{aKMC2}).

\section{Algorithms}
\label{Algorithm}

\subsection{Outline of the main KMC algorithm}
\label{mainAlgorithm}
\begin{figure}[htb!]
\centering
\scalebox{0.5}{\includegraphics{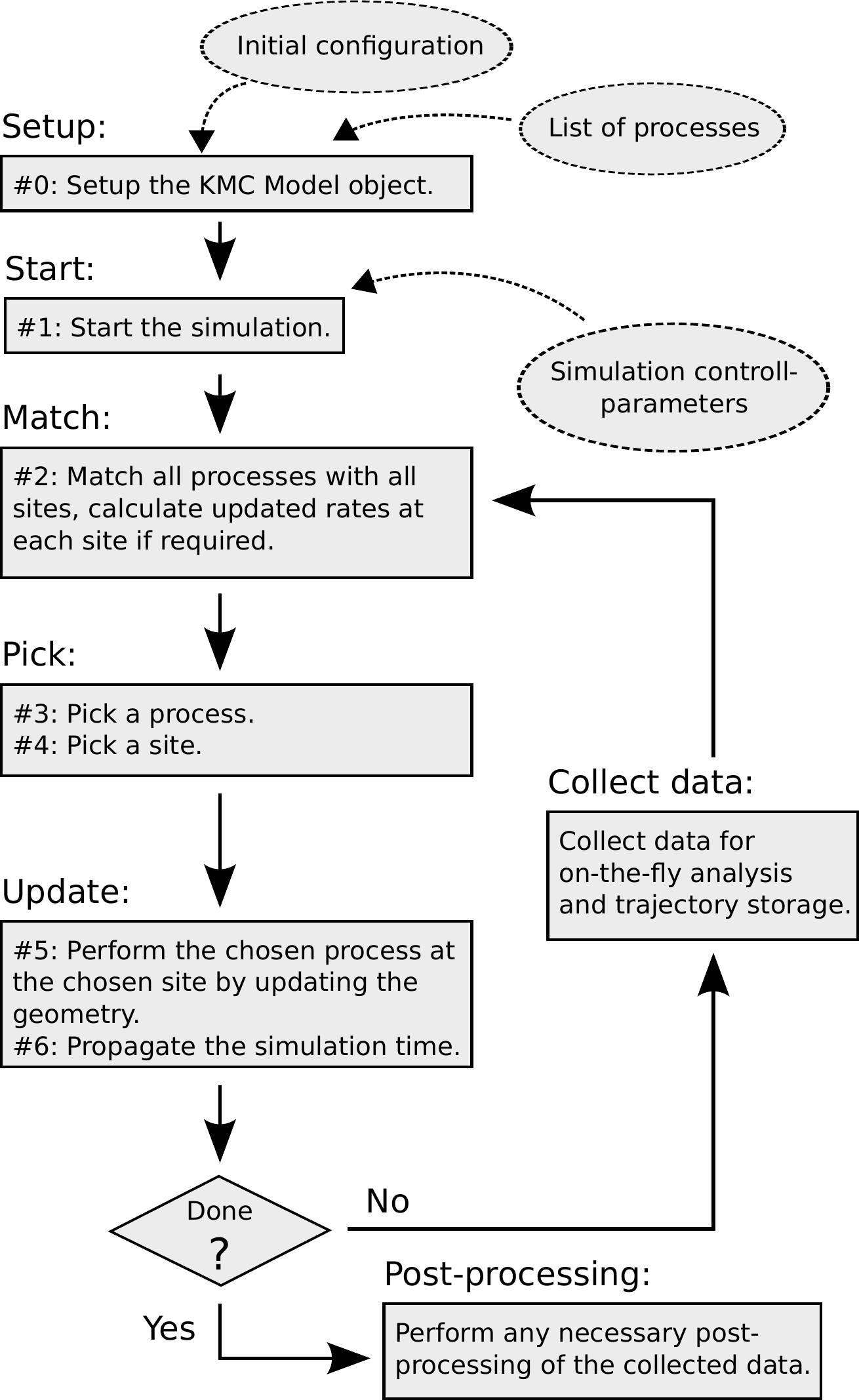}}
\caption{Flowchart of the main KMC algorithm implemented in
  KMCLib. See text for details.}
\label{figAlgorithm}
\end{figure}

Figure \ref{figAlgorithm} shows a schematic illustration of the
version of the VSSM or {\it n-fold} way {\it lattice} KMC algorithm
implemented in KMCLib.

\begin{enumerate}

\item
Start with a set of particles at a set of points in
space (lattice sites), and a list of all possible elementary
processes {\it e.g.\ }diffusion or reaction events that can take the system of
particles from the present state to the next state.
Each elementary process $p_i$ is associated with a rate
constant $r_i$.

\item
Match all processes with the geometry at each lattice site to construct a
table of the availability of each process and obtain the total rate of each
process $R_i$ by multiplying the process' rate $r_i$ with the number of sites
where the process is available $N_i$
\begin{equation}
R_i = r_i N_i
\end{equation}
For the list of processes a vector ${\bf P}$ is constructed with the incremental
total rates, such that for the $j$:th process the incremental value is
\begin{equation}
\label{Pk}
P_j = \sum_{i=1}^j R_i
\end{equation}

\item
A process to perform is chosen at random, proportionally to the
relative total rate of the process. This is done
by drawing a random number $\rho$ between 0 and the last value in ${\bf
  P}$, $P_k$ and a binary search is performed over the elements in ${\bf
  P}$ to determine which process to take. Starting from the lower
end, the last process $p_j$ for which $P_j < \rho$ is chosen.

\item
A site to perform the process at is chosen by drawing from
the list of sites where the process is available, with each
site being equally likely.

\item
The chosen process is performed at the chosen site
by updating the geometry according to the specifications of the
particular process. Each process holds information about the local geometry
around a site before and after the process is performed.
The {\it before} geometry is used for
matching the local geometry of a lattice site with the process, while
the {\it after} geometry is
used to update the local geometry around the lattice site where
the process is performed.

\item
Time is propagated by drawing a random number between 0 and
1, and the time step $\Delta t$ to add to the simulation time is
determined by
\begin{equation}
\label{eqnTime}
\Delta t = \frac{-\ln(\rho)}{P_k}
\end{equation}
with $P_k$ being the total available rate in the system.

\item
Repeat from step 2 until the maximum allowed number of steps is reached.

\end{enumerate}
The simulation is run for a fixed number of steps. Geometry and
time step data can either be stored in a trajectory file for
post-processing and analysis, or analysis can be performed on-the-fly
during the simulation.

\subsection{Algorithm modifications due to custom rate calculations}
\label{modAlgorithm}
An important feature of KMCLib is its ability to handle
processes where the rates are not given at the start of the simulation,
but are instead calculated and updated on-the-fly during the simulation.
The two different ways to determine the process rates (at simulation start-up
or during the simulation) can furthermore be combined, such that some
of the processes gets their rates updated during the simulation while
others do not. The way the rate update framework is implemented allows
for great flexibility allowing the user to define the
desired behavior via
Python code. It is {\it e.g.\ }possible
to couple the update of the process rates
to global or local geometrical parameters, such that a
temperature or pressure gradient can affect the local rates of
diffusion processes differently at different sites, or that the
process rates are being modified at sites in the
vicinity of certain atom types or geometrical features that may arise
during the simulation. The handling of custom rates requires some
modifications to the above described algorithm:

\begin{enumerate}

\item
At start-up, not only are the initial set of particles and their
geometry and
all possible processes given, but also a custom rate
calculator. The custom rate calculator is a Python class that can receive
information about a process and the local geometry around a lattice
site and return an updated rate constant for the process at that
particular site. The custom rate calculator has to follow a well defined API to
assure compatibility with the program and is described in more detail
in section \ref{customRate} below, as well as in the program user manual.

\item
All processes are matched with the geometry at each site to construct a
table of the availability and total rate of each process. At each site where a
process is possible the rate is also updated using the custom rate
calculator. The total rate of each process $R_i$ is now obtained by
summing the updated local rate for each site where the process is
available. Note that due to the use of custom rates the same
elementary process will now have a different rate at different sites.
With $N_i$ being the available sites for process $i$, and $r_j$ being the locally
updated rate at site $j$ the total rate for process $i$ is given by
\begin{equation}
R_i = \sum_{j=1}^{N_i} r_j
\end{equation}
The vector ${\bf P}$ with the incremental
total rate for all processes is constructed according to (\ref{Pk}).
A similar vector ${\bf S}$ of incremental rates is also
constructed for each process with the available sites and their
respective rates
\begin{equation}
\label{Sk}
S_j = \sum_{i=1}^j r_i
\end{equation}
where $r_i$ is the rate at site $i$ of the process.

\item
A process to perform is chosen at random, proportionally to the
relative total rate of the process. This is done in the same way as
described previously by drawing a random number between 0 and $P_k$
and performing a binary search over the ${\bf P}$ vector.

\item
A site to perform the chosen process at, is chosen by drawing from
the list of sites where the chosen process is available. Since the
rate now may vary over the sites at which the process is available it
becomes necessary to draw the site with a similar procedure as when
the process is chosen. A random number $\rho$ is generated between
0 and the last value in ${\bf S}$, $S_j$, and a binary search is
performed over the elements in ${\bf S}$ to determine which site to
take. Starting from the lower end, the last site $i$ for which $S_i < \rho$ is chosen.

\end{enumerate}
The reminder of the algorithm (perform the process, propagate
simulation time and collect data)
is identical with the procedure described in section \ref{mainAlgorithm} above.

\subsection{Outline of the mean square displacement algorithm}
To collect mean square displacement (MSD) data on the fly during the
simulation a history buffer is used where a fixed number of
coordinates and corresponding simulation times are stored for each
particle of interest. Squared displacements of the coordinates are
calculated and stored in a histogram.

\begin{enumerate}

\item
Before the
simulation starts the history buffer is populated with the initial
coordinate and time for each particle.

\item
When a particle is moved its new coordinates along with the current
time of the simulation is stored at the first position in the history
buffer for that particle. Space is made for the new data in the buffer
by moving each coordinate and time already in the buffer to the next
position in memory. If the buffer is full the last value is thrown
away.

\item
The squared displacements between the first (latest added) coordinate in
the buffer and all subsequent coordinates stored are calculated and
binned in a squared-displacements histogram. The number of values added
to each bin is collected in a MSD bin count histogram.

\item
After the simulation is finished the MSD data is obtained by dividing
the value at each bin in the squared-displacements histogram
with the corresponding bin value in the MSD bin
count histogram and the standard deviation of the final MSD data is
estimated.

\end{enumerate}
A more detailed account of the algorithm as well as the
somewhat involved procedure for deriving the correct error
estimate of the final MSD data has been described in detail elsewhere
\cite{MSDAlgorithm}.

\section{Description of the program}
\label{Description}
KMCLib is a general platform from which custom {\it lattice} KMC models can be
implemented.
The Python programming language is used
to define a well tested and well documented user-interface that is
flexible enough to easily define custom KMC models. This is
combined with a well-structured performance
optimized core program written in C++.
The potential need for customization of
the underlying C++ code has largely been eliminated through the use of a
custom rate calculator interface that allows for algorithmic
modifications of the rate determination without the need to
re-compile the underlying C++ code. Analysis can be performed and
data collected during the simulation, with custom analysis modules
loosely coupled to KMCLib via an analysis API in Python.

\begin{figure}[htb!]
\centering
\scalebox{0.4}{\includegraphics{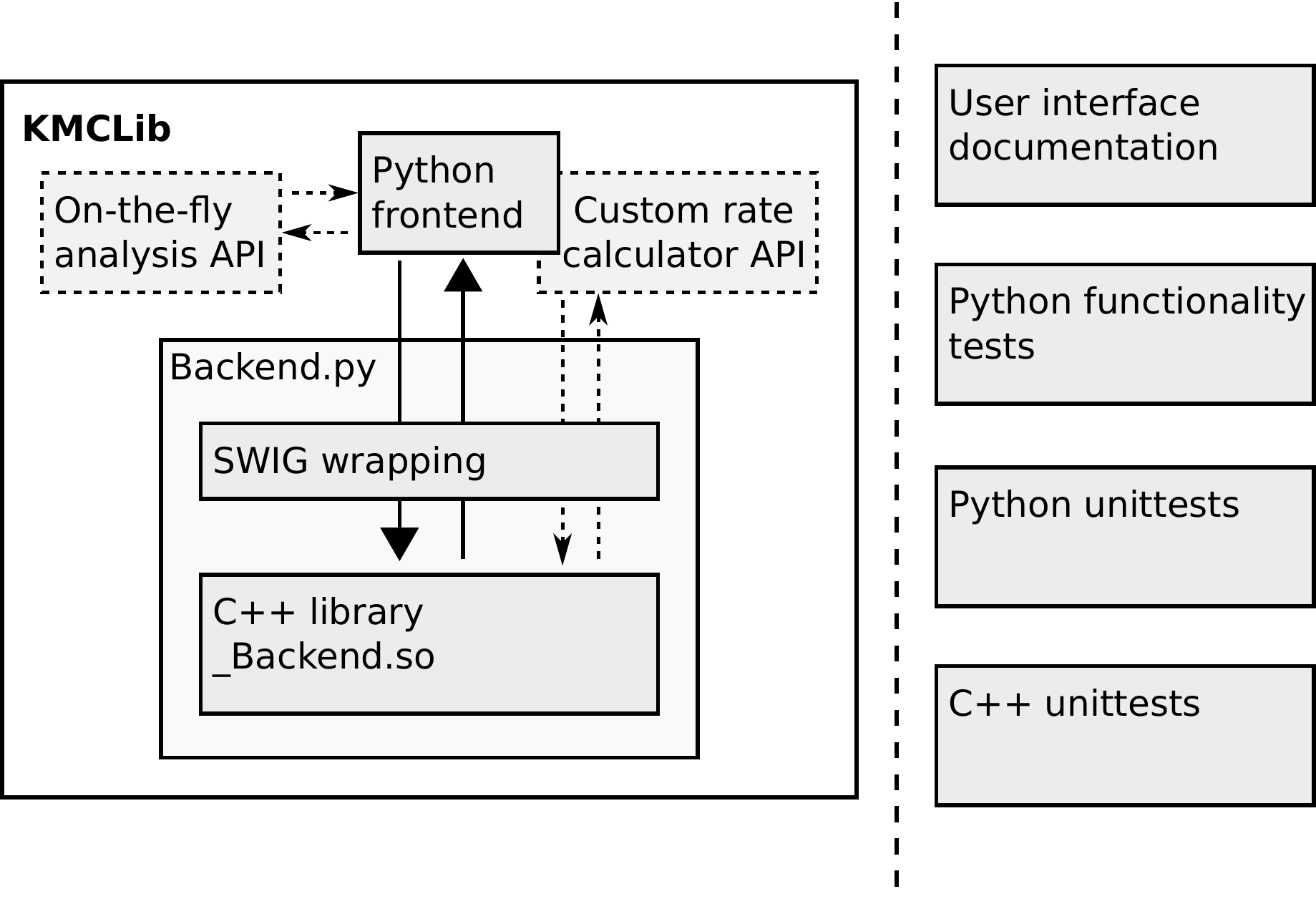}}
\caption{Code organization overview. Deployment code to the left and
  documentation and test code to the right.
  Boxes represent logical entities while arrows
  indicate paths of communication.}
\label{figCode}
\end{figure}

\subsection{Code overview}
Figure \ref{figCode} shows a schematic picture of how the code is
organized. The code consists of two major parts. One part is
the deployment code for actually performing KMC simulations (figure
\ref{figCode} left), while the other part consists of code for
testing and includes also the documentation of the Python
user-interface (figure \ref{figCode} right).

The deployment code consists of a set of Python classes that defines
the user-interface to the {\tt KMCLib} module (including the custom rate calculator API and
the on-the-fly analysis API), as well as a C++ backend library wrapped
to Python using the SWIG framework.
The python frontend is responsible for
handling user-input and all other communication with the outside world,
as well as controlling the flow of the main Monte Carlo loop and
communicating with the C++ backend via the SWIG interface. All
computationally heavy tasks, from step \#2 through to step \#5 in
figure \ref{figAlgorithm}, including matching, picking and updating,
are implemented in the C++ backend.

All software needs testing to ensure its correct
functioning. If the tests are performed systematically and at all levels
of the code the risk of having a malfunctioning program can be
greatly reduced \cite{TDD}. The automated KMCLib tests are split up in three parts: {\it
  functionality tests},
written using the Python {\tt unittest} module testing the whole
program at an aggregate level; {\it Python unittests} (also written
using the Python {\tt unittest} module) testing the Python code at a
detailed level; and the {\it C++ unittests} implementing detailed tests
of the C++ code.
The C++ unittests are setup using the CPPUNIT
framework.
The unittests are intended to test each building block of the code at as low
level as possible, while the functionality tests set up realistic but small simulations using the KMCLib
Python interface much in the same way a user would do.
The functionality tests represent good usage examples
and can as such be used by new users to get their first simulations up
and running.

The Python code is marked up with doc-strings using the Sphinx
documentation format. A manual for the Python interface
can be auto generated from the source code and the
doc-strings using Sphinx and the provided makefile.
The documentation included with KMCLib also describes the installation
procedure and the code usage. All header files in C++ are documented
using the doxygen format
to aid in any deeper level
customization and development work.

\subsection{Input structure}
Input to KMCLib is written as a Python
script and executed using a standard Python interpreter. The input
script should start with loading the KMCLib module using a
{\tt from KMCLib import *}
statement. This will load all API classes that are needed to setup and
run a lattice KMC calculation with KMCLib. Input is then built up by
specifying the geometry and all processes via the {\tt KMCConfiguration}
and {\tt KMCInteractions} classes (see the reference manual for details).

On-the-fly custom rate calculations can at this stage be introduced simply by providing an
instance of a user-given class inheriting from the {\tt
  KMCRateCalculatorPlugin} interface when setting up the
{\tt KMCInteractions} object. If provided, the matching
procedure later in the algorithm will use this calculator to determine
an updated rate for
each process at each site where the process is available.
The {\tt KMCRateCalculatorPlugin} Python class in turn inherits from the
{\tt RateCalculator} C++ class. This is achieved by declaring the C++ base class
a ``director'' in SWIG. When a process is
matched the matching routine in C++ calls the user-specified Python
{\tt CustomRateCalculator} via its inherited C++ interface.
The custom rates interface thus allows the user to provide an altered
rate for the process at the specific site in question. The
capacity to use custom rates greatly
increases the flexibility of the program and opens up the
possibilities for performing KMC simulations with site specific rates
or to take into account longer-range interactions when
determining the rate of each elementary process. Note that the
user is only required to write a Python extension to the
program and there is no need to re-compile the underlying C++ code.
Details of the custom rate calculator interface can be found in the reference manual.

After specifying the geometry and all interactions the input
script sets up an instance of the main simulation object the {\tt
  KMCModel},  and the model is run by calling the {\tt run}
method on the {\tt KMCModel} object with an instance of the {\tt
  KMCControlParameters} class as input holding all information on the
number of steps in the simulation and when analysis and
trajectory data should be gathered.
If on-the-fly analysis is required a list of user-specified analysis
classes, each inheriting from the {\tt KMCAnalysisPlugin} class,
is given as an argument to the {\tt run} method.
Details of how to set up the
necessary objects with the correct input can be found in the
program reference manual.

\subsection{Features and implementation details}

\subsubsection{Match and update algorithm}
Since the matching of local geometries with possible processes plays
a central role in the performance of any lattice KMC
implementation we will below describe our matching algorithm in some detail.

The KMCLib backend uses a type of neighbor lists that we will refer to as
{\it match-lists}. They are used for representing local geometries
around lattice sites in the configuration. When the underlying C++
configuration is constructed from the user-given Python input a
match-list is generated for
each lattice site. Each match-list holds the lattice-site type
information (used for labeling different elements or particle types in the
simulation), the corresponding lattice site indices and the relative local
coordinates for all neighboring lattices sites.
Each lattice site match-lists is sorted according
to its relative coordinates. The size of the match-lists is determined
from the user-defined processes. The process with the largest
included neighborhood will determine the cutoff used for setting up
all lattice site match-lists.

KMCLib has a dual representation of the geometry. The configuration is
represented on a lattice and there is information on each
lattice-point about the lattice-point type. But apart from the lattice
geometry we also have a list of atom-ID tags with their corresponding
current types and Cartesian coordinates. The two representations are connected by the
lattice that keeps track of which atom that sits on each lattice-point.
This might seem trivial at first, but
provides a powerful tool for keeping track of the individual particles
on the lattice, which in turn is a prerequisite for any efficient
diffusion calculation.

Each user given process holds information about the local
configuration around a lattice site before and after the process is
performed, as well as information about how each of the involved
particles should be moved from the initial to the final state of the process.
The local geometry information is stored in match-lists similar to the ones used at the
lattice-sites. We will refer to these match-lists as the {\it before}
and {\it after} process match-lists. The {\it before} process
match-list will be used for matching against local geometries on the
lattice, while the {\it after} process match-list will be used to
update the lattice when the process is performed. Each process
match-list is sorted in the same way as the lattice site
match-lists. If the user has allowed match-list wild-cards
the coordinates of the process match-list are compared to the
corresponding set of local coordinates at the lattice and any missing
entry in the process match-list is inserted with a wild-card
identifier as the type.
The movements of the individual
particles are represented as a set of move-vectors, which in turn will
be applied to the Cartesian coordinates of the atom-IDs involved in
the process.

The matching of a process with a lattice site is performed by comparing the site
match-list with the {\it before} process match-list. Since the lists
are sorted in the same way only the type information needs to be
compared. If wild-cards are not allowed in the process match-lists the
coordinates must also be compared.
Performing a process is done by comparing the site match-list
and the process {\it after} match-list. Any mismatch between the
two will result in the corresponding types information being updated
on the lattice.
The atom-IDs are updated on
the lattice sites according to the specifications given from the
process move-vectors and the move-vectors are furthermore added to the
corresponding atom-ID coordinates. In this way the lattice
configuration and the atom-ID coordinates are guaranteed to be
synchronized, with the important difference that the atom-ID coordinates are
never checked for periodic boundary conditions.

Before the simulation enters the Monte Carlo loop the matching must be
done between all sites and all processes. This results in an $O(n_L
\times n_P \times m)$ scaling behavior, where $n_L$ is the number of
lattice sites, $n_P$ is the number of processes and $m$ is the average
number of match-list entries that is looped over for each match.
After a process has been performed, however,
there is no need to re-match the processes at all sites. Only sites
that are in the vicinity of the site where the process was performed
are affected by the move and a number of affected sites $n_A \ll
n_L$, which is constant with the size of the simulation will require a re-match.
This results in an $O(n_A \times n_P \times m)$ scaling behavior of
the matching in the Monte Carlo loop, and the matching is thus
{\it constant} $O(1)$ with respect to the size of the simulation.

\subsubsection{On-the-fly rate calculations}
\label{customRate}
To use the custom rate calculation abilities with KMCLib it is enough
to define a class that inherits from {\tt CustomRateCalculator}, and
overload its {\tt rate} method. The custom rate calculator is
given to the interactions object using its {\tt setRateCalculator}
method before the interactions object is used to construct the KMC model.

When a rate update is required for a process at a specific site the
custom rate calculator's {\tt rate} method will be called with
parameters specifying the global coordinates of the site
and its local geometry and elements (types) before and after the process is
applied. When the {\tt rate} method is called with a site and a
process it is guaranteed that the process is available at the particular
site (i.e.\ that the site and the process match).
The rate constant associated with the
process is also given and it is then fully up to the user to specify
any modifications to this rate from the given geometry and process
input. Details of the custom rate interface
can be found in the reference manual.

It should be noted that the flexibility of using custom rate updates
inevitably comes with a performance penalty. It is therefore
not generally recommended to implement models using a custom rate
calculator that can easily be implemented with processes with fixed
rates. As an illustrating example we can look the Ising model of
\cite{OriginalKMC1} that is provided with the functionality tests, where it is
implemented both using a custom rate calculator in Python and using processes with
fixed rates. The equivalent simulation with fixed rates runs roughly
four times faster.

\subsubsection{Trajectory format}
A simulation trajectory can be produced in two formats using KMCLib,
depending on the type of simulation. For pure lattice KMC simulations
that are only concerned with lattice sites and their types, without keeping
track of the movement of specific atoms, a lattice trajectory format is
used where the type at each lattice site is stored. Since the lattice
positions are fixed during the simulation it is enough to print them
only once to the trajectory file.

For simulations where move-vectors have been provided with
the processes, so that the program can keep track of the individual
atoms during the simulation, it is also possible to use a simple xyz
format that prints the type and coordinate for each atom along the
simulation. Note that although the lattice sites do not move during
the simulation the individual atoms do move according to the provided
move vectors.

The user can set the interval between steps to save to the
trajectory. To avoid unnecessary file IO operations the trajectory
data is stored in an internal buffer. The buffer size and time since
last file IO is saved and the trajectory buffer is printed to file
only when the buffer has reached its size limit or when a
long enough time has passed since last print to file.

\subsubsection{On-the-fly analysis}
\label{analysis}
Instead of storing trajectory data to file it can
be handy to run analysis directly while running the simulation. KMCLib
provides the possibility to perform on-the-fly analysis via
an analysis plugin interface. The user defines an object that inherits from
the {\tt KMCAnalysisPlugin } class and overloads the {\tt setup}, {\tt
  registerStep} and {\tt finalize} methods for desired behavior.

Right before the Monte Carlo loop is entered each user-provided
analysis object will in turn get access to the initialized
system. This is done by calling each analysis object's {\tt
  setup} method with the configuration and simulation start
time as arguments, allowing for the analysis objects to set up
internal memory and collect initial data.
After a step in the Monte Carlo loop has been performed program flow
control is returned to Python where all required analysis and trajectory
handling is performed (the ``collect data'' step in figure \ref{figAlgorithm}).
Analysis is performed by letting each user-provided
analysis object in turn get access to all details of the present state
of the simulation by calling each analysis object's {\tt registerStep} method
with the configuration and simulation time as arguments. It is then up
to each analysis object to extract relevant information at this stage
in the simulation.
At the end of the simulation the {\tt finalize} method is called on
each analysis object to allow for post-processing of collected data
before the program stops.

The mean square displacement analysis described in section \ref{MSD}
below is implemented using the analysis plugin interface. A detailed
account of the analysis plugin interface can be found in the reference manual.

\subsubsection{On-the-fly mean square displacements}
\label{MSD}
To facilitate diffusion studies with KMCLib the
algorithm described in \cite{MSDAlgorithm} for
calculating mean square displacements with correct error estimates
on-the-fly during a simulation is implemented.
To run a simulation with mean square displacement analysis it is
required that move vectors are provided with the processes to allow the
program to keep track of the individual atoms. An atom type to track
must be provided by the user as well as parameters specifying the
number of history steps to use and the histogram spacing. It is also
important that the primitive unit cell of the configuration is given
in correct units, since the transformation to Cartesian coordinates is
done using this information before summing contributions for each time-step.

The mean square displacement analysis is implemented as an analysis
plugin as described in section \ref{analysis} above and
several examples of its use are included in the functionality tests.

\subsection{Parallelization strategy}
\label{parallel}
The MPI parallelization scheme in KMCLib is based on the
assumption that the matching of a process with a local geometry,
including rate evaluation when required, is by far the most
time-consuming part of each step in the Monte Carlo loop. This
condition is particularly well satisfied for all cases where
custom rate calculations involve anything more than a simple table look-up.

When the list of sites to match against the list of processes is
generated after a process is performed,
the process-site pairs to match are distributed over all ranks.
The matching is performed in parallel and the
matching result is communicated to all ranks. If a custom rate calculator
is used the list of process-site pairs that need their rates to be updated
is also split up over all ranks and calculated in
parallel, and the updated rates are communicated to all ranks.
This parallelization strategy is similar to
the approach in aKMC \cite{aKMC2} where the expensive saddle-point
searches are performed in parallel. Contrary to parallelization schemes
that rely on a spatial decomposition of the simulation box with
several processes performed simultaneously on different ranks,
our approach does not affect the overall serial algorithm. Running in
serial or parallel is thus guaranteed to generate exactly the same
simulation results.

To achieve a good work-load using this parallelization scheme it is necessary
that the number of process-site pairs to parallelize over is much
longer than, or exactly matches, the number of MPI ranks.
As will be apparent in section \ref{ParallelPerformance} below,
the parallelization will furthermore, as expected, work more efficiently for
simulations with heavier custom rate calculations.

\subsection{Random numbers}
Good quality random numbers are absolutely crucial for results from
any Monte Carlo method to be valid \cite{AllenTildesley}. Over the
years, the quality of available pseudo random numbers has improved
significantly and today the Mersenne-Twister \cite{Mersenne-Twister}
algorithm is considered the gold
standard for non-cryptographic applications.
A publicly available library Mersenne-Twister
implementation \cite{LarsenMT} is therefore included in KMCLib.

To assure that the included random number generator works correctly
we tested it against the standard C++11 implementation provided with
the gcc (Ubuntu/Linaro 4.6.3-1ubuntu5) 4.6.3 compiler.
Both pseudo random number generators generated identical results
for several million consecutive numbers.
Interchanging the included random number generator with
the standard C++11 implementation in KMCLib furthermore
produced identical simulation results.
An internal wrapper was finally implemented around the pseudo random number
generator in KMCLib to facilitate use of any other pseudo random
number generator implementation, or the standard C++11 implementation,
with minimal modifications to the code.

\section{Timings and scaling}

We use two different systems for demonstrating the scaling behavior of
KMCLib.
The first system is a simple one-dimensional A-B-C model, where
particles of type A, B or C sit on a one-dimensional lattice. Only two processes
are used for this system, the flipping of A to B and B to A, leaving
the C particles inert. This allows us to vary the total size of the
system while keeping the number of active A+B sites constant, simply
by varying the number of C particles. Both
included processes have the same rate and
the number of A particles was always the same as
the number of B particles at the start of each simulation.

The second system is a more complex model of oxygen-vacancy diffusion in
a fluorite metal oxide structure with a large fraction of dopants on
the metal sites and vacancies in the oxygen sub-lattice.
The primitive cell is based on the cubic
oxygen sub-lattice. There are two basis points in the cell, (0,0,0)
and (0.5,0.5,0.5). Since we use the primitive cell of the oxygen
sub-lattice to represent a fluorite metal oxide structure half the (0.5,0.5,0.5) sites will be occupied with metal
ions and half will be marked as empty.
Oxygen-vacancy diffusion was described with six
processes describing nearest-neighbor hops along the Cartesian
axes. The system was run with two different settings. In the first
setting oxygen diffusion was modeled with equal rates for all six processes. In
the second setting the system was run with a custom rate calculator that
modified the process rates based on the local distribution of
dopants within three shells of primitive cells (258 neighbor sites)
from the vacancy site of interest.

\begin{figure}[thb!]
\centering
\scalebox{0.5}{\includegraphics{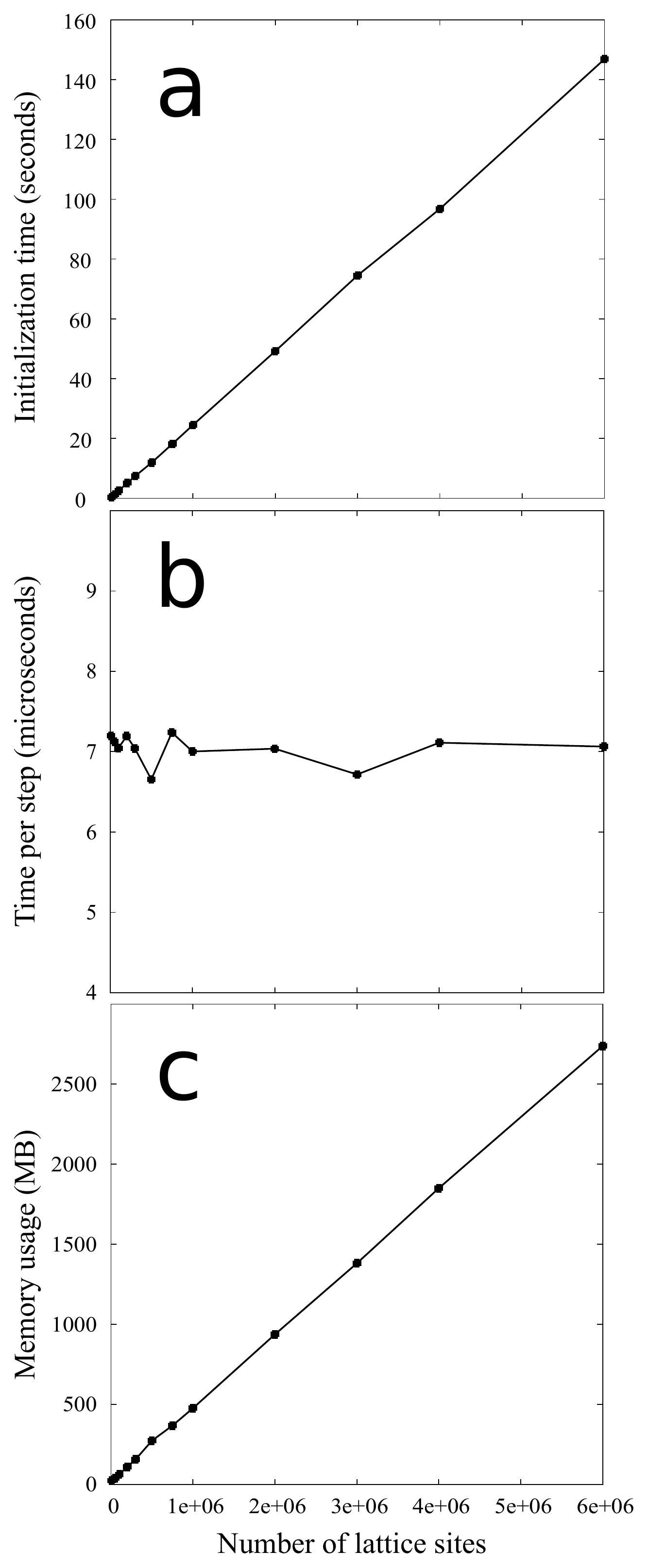}}
\caption{Scaling behavior (A-B-C model) when varying the total
  number of lattice sites while keeping the number of active (A+B)
  sites constant. The initialization time (a) and memory usage (c)
  grows linearly with the system size, while the time per step in the
  loop (b) stays constant.}
\label{figABC}
\end{figure}

\begin{figure}[thb!]
\centering
\scalebox{0.5}{\includegraphics{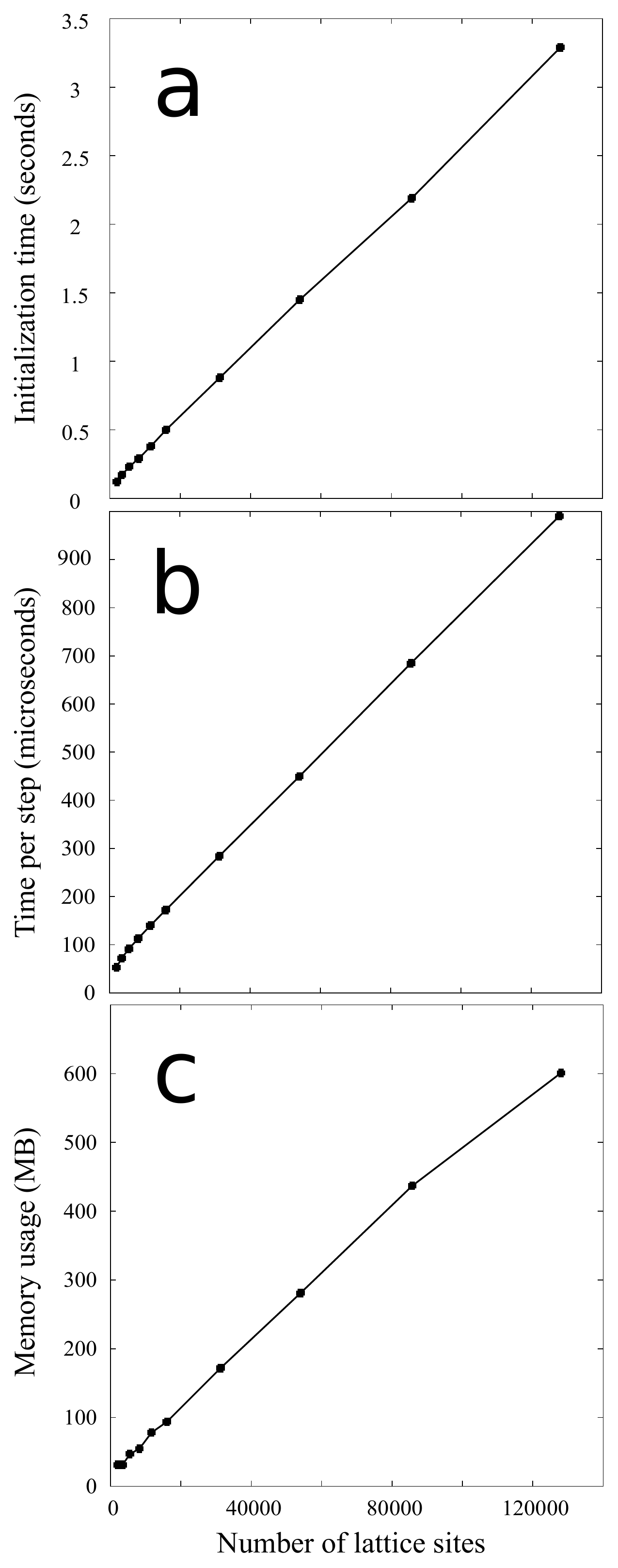}}
\caption{The scaling behavior of the fluorite metal oxide model in the
  first setting (with short rate evaluation time, see text).
  The system size is varied keeping the ratio
  between different lattice site types constant. The
  initialization time (a) and memory usage (c) grows linearly
  with system size. The short time spent in rate evaluation makes
  updating the lists of possible sites for each process the dominating
  term and the time per step in the loop (b) therefore grows
  linearly with system size.}
\label{figFluorite1}
\end{figure}

\begin{figure}[htb!]
\centering
\scalebox{0.5}{\includegraphics{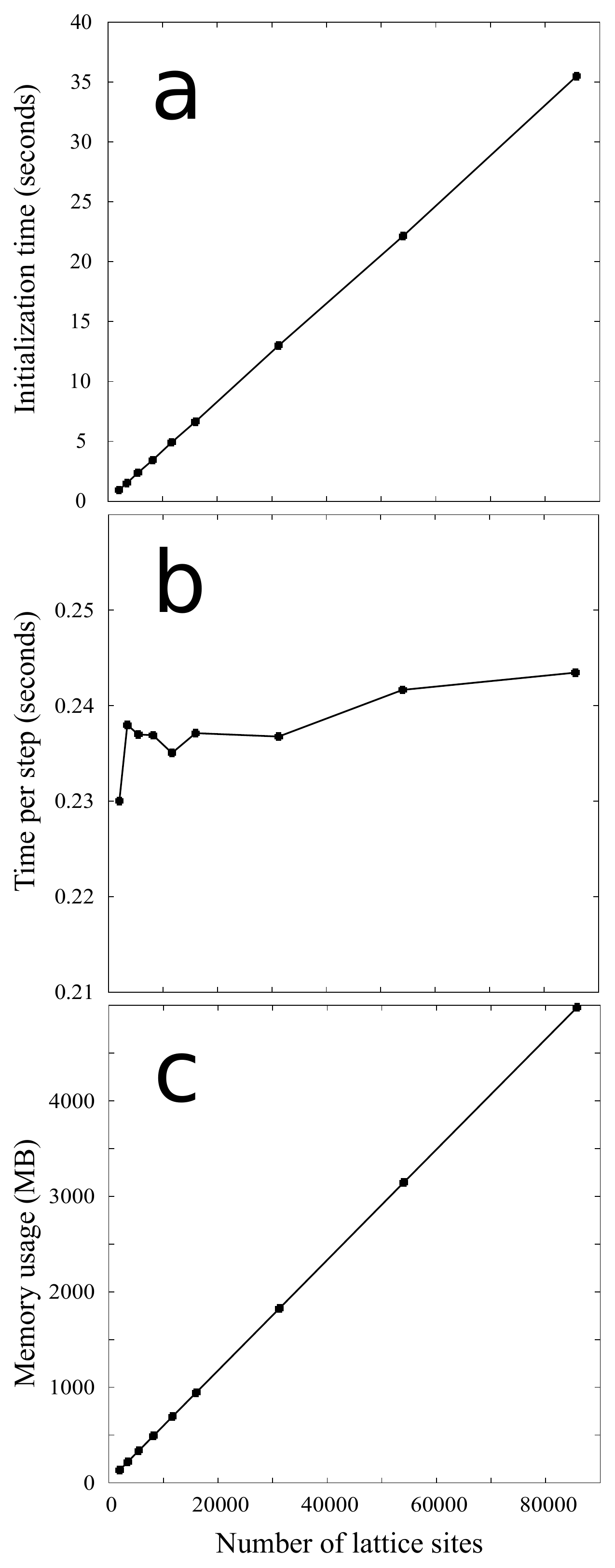}}
\caption{The scaling behavior of the fluorite metal oxide model in the
  second setting (with long rate evaluation time, see text).
  The system size is varied with the ratio
  between lattice site type occupation kept constant. The
  initialization time (a) and memory usage (c) grows linearly
  with system size. The rate evaluation is here the dominating term
  and the time per step in the loop (b) is therefore close to constant.}
\label{figFluorite2}
\end{figure}

\subsection{Serial performance}
All serial performance tests were carried out on a laptop computer running
Ubuntu-12.04 LTS, with an Intel i7-3517U 1.90 GHz CPU with 4 MB cache,
and with 8 GB RAM. Only one of the CPU's cores was used.

In figure \ref{figABC} we show the scaling with system size for the
simple A-B-C model when the number of active (A+B) sites are kept
constant at 2000, varying only the number of C sites. The
initialization time (figure \ref{figABC}a) is seen to grow linearly
with the system size. This is because the initialization loops over
all sites to set up their match-lists. Linearity is achieved by
dividing the simulation box up in blocks of primitive unit-cells and
restricting the calculation of distances for the match-lists within a
limited number of such neighboring cells. To make sure all relevant
neighbor information is included the number of neighbor cells
to consider is determined by the longest process cutoff as defined by
the user. The memory usage (figure \ref{figABC}c) is also seen to grow
linearly with system size as expected.
The time per step in the loop (figure \ref{figABC}b) is constant. This
is also expected since no extra work should be required per step as long as
the number of active (A+B) sites is kept constant.

Figure \ref{figFluorite1} shows the scaling with system size for the
fluorite metal oxide oxygen-vacancy diffusion system in the first
setting. Both the initialization time and memory usage
(\ref{figFluorite1}a and c) scales linearly with system size as
expected. Contrary to the constant scaling for the A-B-C model, the
time spent for each step in the loop (figure \ref{figFluorite1}b)
also grows linearly with system size. This linear behavior arises since
the system size is scaled homogeneously so that the number of active
sites (sites where the processes can be applied) grows linearly with
the total size of the system. When a site is matched against a process
the list of available sites for that process must be searched through
for the site of interest to determine if the site should be added to
(or removed from) this list. When the number of available sites grows
with system size the list to search through grows accordingly, which
gives rise to the observed linear scaling behavior.

The situation is quite different for the fluorite metal oxide
oxygen-vacancy diffusion system in the second setting
(figure \ref{figFluorite2}). While the
initialization time and memory usage (figure \ref{figFluorite2}a and c) still
scales linearly, the time per step in the loop
(figure \ref{figFluorite2}b) is close to
constant, with only a slight increase with system size. This behavior
is expected and can be explained by noting the difference in scale on
the y-axis in figures \ref{figFluorite1}b and \ref{figFluorite2}b. The
linear component from figure \ref{figFluorite1}b is still present in
figure \ref{figFluorite2}b, but the time spent in the loop is
completely dominated by the rate evaluation resulting in close to
constant $O(1)$ scaling.

We expect this type of roughly constant $O(1)$ scaling for most realistic
applications even when custom rates are not used, since the matching
step is typically far more time consuming than the search through
the availability vectors of the processes.

\subsection{System size and simulation time}
A peculiarity when it comes to the scaling behavior of any KMC
simulation implementing the VSSM algorithm,
which at first can be easy to overlook and therefore well worth
noting, is how the simulation time (as opposed to the wall-clock time
measured in the timings above) varies with the system size. We recall
that the simulation time is propagated according to the expression
(\ref{eqnTime}) where
the sum of the rates in the system appears in the denominator. The
length of each time-step in the simulation is thus inversely
proportional to the total rate in the system. This means that
doubling the size of the system (and the number of available sites)
and running the simulation for the same number of
elementary steps will only propagate the simulation time half as far.

\begin{figure}[htb!]
\centering
\scalebox{0.6}{\includegraphics{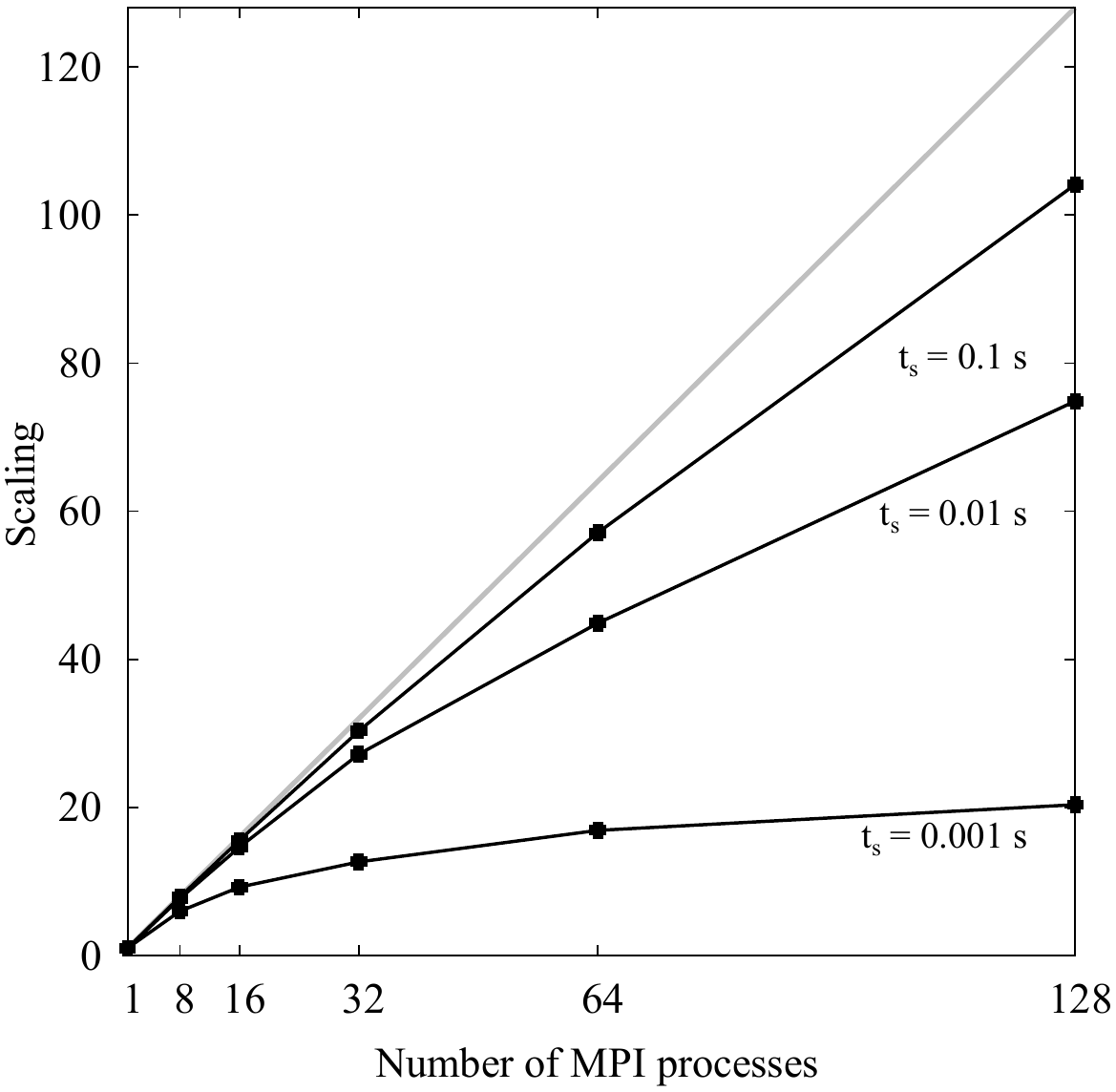}}
\caption{Scaling with the number of MPI processes for three cases
  with different time spent in the custom rate calculation (0.001,
  0.01 and 0.1 seconds additional time per rate evaluation as
  indicated). Black dots represent measured values
  and the gray line indicates ideal linear scaling.}
\label{figMPI}
\end{figure}

\subsection{Parallel performance}
\label{ParallelPerformance}
All parallel performance tests were run on a CentOS Linux-based
cluster with HP Proliant SL2x170z nodes, each equipped with two
quad-core Intel Xeon E5520 CPUs and 36 GB RAM, with infiniband
interconnect. One core was used for each MPI process.

As discussed in section \ref{parallel} above the parallel performance is
expected to depend strongly on the amount of work needed
during the custom rate calculations. To run the same test system
for different times spend in the rate calculation function a sleep
statement was introduced. Three different cases were considered,
waiting an additional 1, 10 and 100 milliseconds respectively each
time the rate calculation function was called.

Figure \ref{figMPI} shows the scaling calculated as the total time
running on one MPI process over the total time running in parallel.
As expected, the
more time spent in the custom rate calculation the better the
scaling. When only 1 millisecond is added to the rate calculation the
scaling flattens out above 16 processes, but with 100 additional milliseconds
in the rate calculation the scaling is excellent all the way up to
the 128 processes.

Clearly, the heavier the custom rate calculations the more is to gain from
running in parallel, but the number of sites to re-match at each step
will also influence the parallel performance.
The test system used in figure \ref{figMPI} is three-dimensional with two
basis sites for each primitive unit cell
and a large cutoff which resulted in more than 500 process-site pairs
to re-evaluate at each step. For a good load balance it is necessary
that the number of process-site pairs to re-match is either much
larger than, or an exact multiple of, the number of MPI processes.
Running efficiently in parallel will therefore be limited to a few
MPI processes for systems with few neighbors (with low dimensionality)
and short cutoff.

\subsection{Caching of calculated rates}
\label{Caching}
In those cases where the on-the-fly rates are kept constant over time,
i.e., when no time-evolution is allowed in the custom rate expressions,
it is possible to use caching techniques to save calculated rates for
later retrieval when needed. The efficiency of such a caching scheme
is highly dependent on the details of the simulated system and on
the caching algorithm used, but it can in some cases provide a
significant speedup. For one of the simpler systems investigated,
the two-dimensional Ising spin model (see section \ref{customRate}),
the custom rate implementation can be sped up almost to the
level of the fixed rates implementation using a prototype caching
scheme where all the calculated custom rates are saved.

For systems with larger number of possible lateral interactions saving
all calculated rates is not feasible due to memory limitations.
A caching scheme with several internal hash tables with fixed maximum
size, where the oldest table filled is emptied when the
latest table reaches its maximum size limit would resolve
the memory issue. No cashing mechanism is implemented in the present
version of the code but will be further investigated for future
releases, however a prototype implementation is available in a development
branch of the KMCLib git repository.
The structure of the custom rate framework meanwhile makes
it easy for the user to implement her own cashing mechanism.

\section{Concluding remarks}

We have presented a general framework for lattice KMC
simulations designed to easily be customized, to facilitate
implementation of custom KMC models without having to re-do all the
work of implementing and testing the simulation engine for each new set of
problems to investigate. We have made our code freely available
since we believe this will be of great benefit for the
larger research community, by saving time and letting researchers focus
their effort on the design of new models to solve
challenging physical problems, rather than spending time re-implementing
a well-known method.

This paper has focused mainly on technical details and on
the modifications we have made to the standard lattice KMC algorithm. We have
demonstrated the scaling behavior in serial and shown how the
parallel performance depends on the simulated system. The code
itself is separately documented from a users
perspective, and includes several usage examples in the functionality
tests. We have therefore not given a full usage example here, but
trust that the reader will find the code manual and examples enough to get
started. To run the tests and set up new calculations require
some knowledge of Python programming but following the examples and
the documentation this barrier should be easy to overcome.

The modular plugin design of the program with a well-defined API
is ideal for sharing programming work with the community. Custom rate
calculators,
analysis modules and scripts
for setting up geometries and processes and to visualize results
can be written and published separately,
and we hope to see many such contributions from the research community,
to include in future releases.

\vspace{10mm}
\noindent
{\bf Acknowledgments }\\\\
We gratefully acknowledge Christian Stigen Larsen for making his
Mersenne-Twister implementation publicly available under the GPL license.
We acknowledge financial support by the Swedish Energy Agency
(Energimyndigheten, STEM), the Swedish Research Council
(Vetenskapsr{\aa}det) and the Carl Trygger Foundation.
Supercomputer time was granted by the Swedish National Infrastructure
for Computing (SNIC).

\bibliographystyle{elsarticle-num}
\bibliography{ref.bib}

\end{document}